\documentstyle[prc,aps,epsf,multicol,graphicx]{revtex}

\begin{document}

\title{Unblocking of the Gamow-Teller strength in stellar electron
capture on neutron-rich Germanium isotopes}

\author{K. Langanke$^1$, E. Kolbe$^2$  and D.J. Dean$^3$}
\address{$^1$Institute of Physics and Astronomy, University of
{\AA}rhus, DK-8000 {\AA}rhus C, Denmark\\
$^2$ Departement f\"ur Physik und Astronomie der Universit\"at Basel,
Basel, Switzerland\\
$^3$ Physics Division, Oak Ridge National Laboratory, Oak Ridge, TN
37831 USA }

\maketitle

\begin{abstract}

We propose a new model to calculate stellar electron capture rates
for neutron-rich nuclei. These nuclei are encountered in the
core-collapse of a massive star. Using the Shell Model Monte Carlo
approach, we first calculate the finite temperature occupation
numbers in the parent nucleus. We then use these occupation numbers
as a starting point for calculations
using the random phase approximation. Using the RPA approach, we
calculate electron capture rates including both allowed and forbidden
transitions. Such a hybrid model is particularly
useful for nuclei with proton numbers $Z<40$ and neutron numbers
$N>40$, where allowed Gamow-Teller transitions are 
only possible due to configuration mixing by the residual interaction and by
thermal unblocking of $pf$-shell single-particle
states. Using the even germanium isotopes
$^{68-76}$Ge as examples, we demonstrate that the
configuration mixing is strong enough to unblock the Gamow-Teller
transitions at all temperatures relevant to core-collapse
supernovae.
\end{abstract}


Current one-dimensional core-collapse supernova models fail to produce
explosions \cite{Janka,Mezzacappa}.
A forefront area of research involves determining whether this
failure is due to incorrect microphysics input or whether
it implies that the explosion actually requires multidimensional
effects like convection and rotation. More likely, both the
microphysics and hydrodynamical effects will play important roles
in enhancing our understanding of supernovae explosions.
One important aspect of microphysics that determines
the fate of a core-collapse supernova is electron capture on protons
and nuclei. These weak captures serve to deleptonize the core of
the massive star and determine the final electron fraction, $Y_e$, and
therefore they set the size of the homologuous core.

In his review \cite{Bethe90}, Bethe described the development of
theories and models of electron-capture in supernovae environments.
Early models assumed that the
capture takes place on free protons. This view was revised by
Bethe and collaborators \cite{BBAL} who noted the low concentration of
free protons relative to iron-group
nuclei and the strong Gamow-Teller
(GT) transitions for $f_{7/2}$ protons to be changed
into $f_{5/2}$ neutrons. These authors concluded that electron
capture during the collapse phase takes place on nuclei
in the mass range $A$=60-80.  Subsequently, Fowler, Fuller,
and Newman developed the formalism for stellar weak processes \cite{FFN} 
and estimated the
rates for electron capture on nuclei with $A \leq 60$ (and for
other weak processes) on the basis of the independent particle model and
available experimental data. Recent
decisive progress in nuclear modeling, coupled with
computational advances, made possible reliable calculations of stellar electron
capture and beta-decay rates. These rates were calculated for
$pf$-shell nuclei in large shell-model spaces \cite{Langanke00}.
Although these improved rates lead to
significant changes in the supernova progenitor models \cite{Heger00,Heger2},
they confirm the FFN results: electron captures in
supernova progenitor models indeed take place on complex nuclei in the
iron mass range.

While the iron region has been sufficiently investigated, we know
that the distinct possibility exists for electron capture to occur in nuclei
beyond the $pf$-shell. Historically, this possibility has not
been included in core-collapse simulations.
Fuller \cite{Fuller82} pointed out that the
Pauli principle blocks Gamow-Teller transitions by neutrons
if one uses the independent-particle model with $Z < 40$ and
$N\ge 40$. In this model, the $pf$-shell is completely occupied
by neutrons. It was concluded that electron capture in this
region once again proceeds by free
protons. Modern collapse simulations
still treat electron capture on the basis of the
independent particle model (even reduced to a model       which only
considers $f_{7/2}$ and $f_{5/2}$ orbitals \cite{Bruenn85})    and
also block all capture on nuclei with $N \ge 40$
\cite{Mezzacappa93}. In contrast, Cooperstein and Wambach noted
from an investigation based on the random phase approximation
\cite{Wambach}, that electron capture on
neutron-rich nuclei with protons in the
$pf$-shell and neutron number $N>40$ can compete with
capture on free protons if one
considers forbidden transitions in addition to allowed ones. They also
demonstrated that at high enough
temperatures, $T \sim 1.5$ MeV, Gamow-Teller transitions are
thermally unblocked primarily as a result of the excitation of
neutrons from the $pf$-shell into the $g_{9/2}$
orbital. This unblocking allows GT transitions within the $pf$-shell,
which then again dominate the electron capture
rates. We will argue in this Letter that electron capture on nuclei with
$N>40$ is also dominated
by GT transitions even at rather low stellar temperatures near $T=0.5$
MeV. This effect occurs since configuration mixing induced by
the residual interactions and thermal excitations are already
strong enough to unblock the GT transitions.

We consider a stellar environment with temperature $T$. The
total cross section for capture of an electron with energy $E_e$
(rest mass plus kinetic)
on a nucleus with charge $Z$ and mass number $A$ is given by
\begin{equation}
\sigma (E_e,T) = \frac{G_w^2}{2 \pi} \sum_{i}
F(Z,E_e) \frac{(2J_i+1) e^{-E_i/(kT)}}{G(Z,A,T)}
\sum_{f,\lambda}
{(E_e-Q_{if})^2} \frac{|\langle i|T_\lambda| f \rangle|^2}{(2J_i+1)}\;,
\end{equation}
where $G_w$ is the weak-interaction coupling constant and $F(Z,E_e)$ is
the Fermi function that accounts for the Coulomb distortion of the
electron wave function near the nucleus (see, for example,
\cite{Langanke00}).  The sum over initial states involves a thermal
average of levels, with excitation energies $E_i$ in
the parent nucleus;
$G(Z,A,T)$ is the respective partition function.          Each
initial state $i$ is connected to various final states $f$ via
multipole operators $T_\lambda$ which were derived
in \cite{Walecka}. These operators, in principle, depend on
the momentum transfer; however, at the energies
involved here, momentum transfer
is small. Under these conditions, the 
$\lambda=1^+$, $T_\lambda$ operator reduces
to the Gamow-Teller operator $\sigma \tau_+$ (which changes a
proton into a neutron). We include the so-called quenching of the
GT strength by multiplying the GT transition matrix element by
the constant   factor $0.7$
\cite{Wildenthal,Langanke95}. The
Q-value for a transition between initial
and final states is given by
$Q_{if} = M_f-M_i+E_f-E_i = Q + E_f - E_i$,
where $M_{i,f}$ are the masses of the parent
and daughter nuclei and $E_f$ is the excitation energy of the final
state.

The nuclei of interest in this study are expected to contribute to the
stellar electron capture rates for temperatures $T \approx 0.5-1.5$ MeV.
At such high temperatures an
explicit state-by-state evaluation of the sums in equation (1) is
impossible with current nuclear models. As was noted and applied in
\cite{FFN,Aufderheide94}, the cross section expression becomes significantly
simplified assuming the Brink hypothesis. Brink conjectured that the
strength distribution of the multipole operators in the daughter nucleus
is the same for all
initial states and shifted by the excitation energy of the initial
state.  By using this approximation, the sum over final states becomes
independent of the initial state and the sum over the Boltzmann weights
cancels the partition function. Although the shell-model calculations
of \cite{Langanke00} in fact verify this approximation for $pf$-shell
nuclei in the temperature
range of interest, it cannot be naively applied to the nuclei
here since the unblocking of the Gamow-Teller strength
should be strongly state-dependent; i.e., the
probability of $g_{9/2}$ configuration mixing
will increase with excitation energy. Cooperstein and Wambach
accounted for this effect by a state-by-state evaluation of the cross
section: they reduced the sum over initial states to the spectrum
of the RPA single-particle levels.  As the energy splitting between the
$g_{9/2}$ orbital and the pf-shell is of the order of 2-3
MeV, the thermal unblocking required quite high
temperatures in
\cite{Wambach}.

At the present time, detailed shell-model calculations for
neutron-rich nuclei with $Z<40$ and $N\ge 40$ are feasible
only for a few nuclei.
Furthermore, the
development of a generally reliable shell-model
interaction for nuclei in this region has not yet occurred.
In order to make some progress in this astrophysically
important region, we therefore propose a
model for the calculation of electron capture on heavy  nuclei that is
computationally feasible and incorporates relevant nuclear  structure
physics. We also assume the Brink hypothesis; however,
we apply it to an initial state  which
represents the thermal average of many-body states in
the parent nucleus at temperature $T$. We describe this thermally
averaged initial state by a Slater determinant with
partial occupation numbers for
the relevant single-particle states. Obviously the partial
occupation numbers are a function of
temperature. Adopting the Brink hypothesis
and describing the initial state by the representative
Slater determinant with temperature-dependent occupation
numbers $n_i(T)$, the
electron capture cross section reduces to
\begin{equation}
\sigma (E_e,T) = \frac{G_w^2}{2 \pi} F(Z,E_e)
\sum_k {(E_e-Q - \omega_k)^2} \sum_\lambda S_\lambda (\omega_k,T)
\end{equation}
where $\omega$ is the excitation energy in the daughter nucleus and
$S_\lambda$ is the discrete RPA response for the multipole operator $\lambda$.
In the RPA approach \cite{Kolbe},  multipole operators up
to $\lambda=2$ are considered, and the single-particle energies are
taken from a standard
Woods-Saxon parametrization that is very close to the values used
in Ref. \cite{Wambach}.

We obtain the thermal
occupation numbers that we use
in the RPA calculations from canonical Shell Model Monte
Carlo (SMMC) \cite{Lang} calculations.
SMMC techniques have been
demonstrated to well describe the thermal properties of
nuclei
\cite{Koonin}. For the germanium isotopes $^{68-76}$Ge, we adopted the
complete $(pfg_{9/2})$ shell-model space and used a pairing+quadrupole
residual
interaction \cite{Bes} with parameters appropriate for this region.
Two reasons guide our choice here: first, a reliable shell-model
interaction for neutron-rich nuclei in this region is not
yet available; second, we wish to avoid the Monte Carlo
sign problem associated with using realistic interactions
in SMMC
\cite{Koonin}. By avoiding the sign problem, we are able to
calculate occupation numbers with less statistical
error. For our purposes, this constitutes
a reasonable first attempt to incorporate the relevant
many-body physics and to understand the
effects of temperature and residual interaction      on the
Gamow-Teller strengths in neutron-rich nuclei.
We adopted the single-particle
energies from the KB3 interaction
\cite{Zuker81}, but we artificially reduced the
$f_{5/2}$ orbital by 1 MeV to simulate the effects of the
$\sigma \tau$ component that is missing in our residual interaction.
We assumed an energy splitting of 3 MeV between the $g_{9/2}$ and the
$f_{5/2}$ orbitals.

\begin{figure}[]
   \begin{center}
 \leavevmode
 \includegraphics[width=0.5\columnwidth,angle=270]{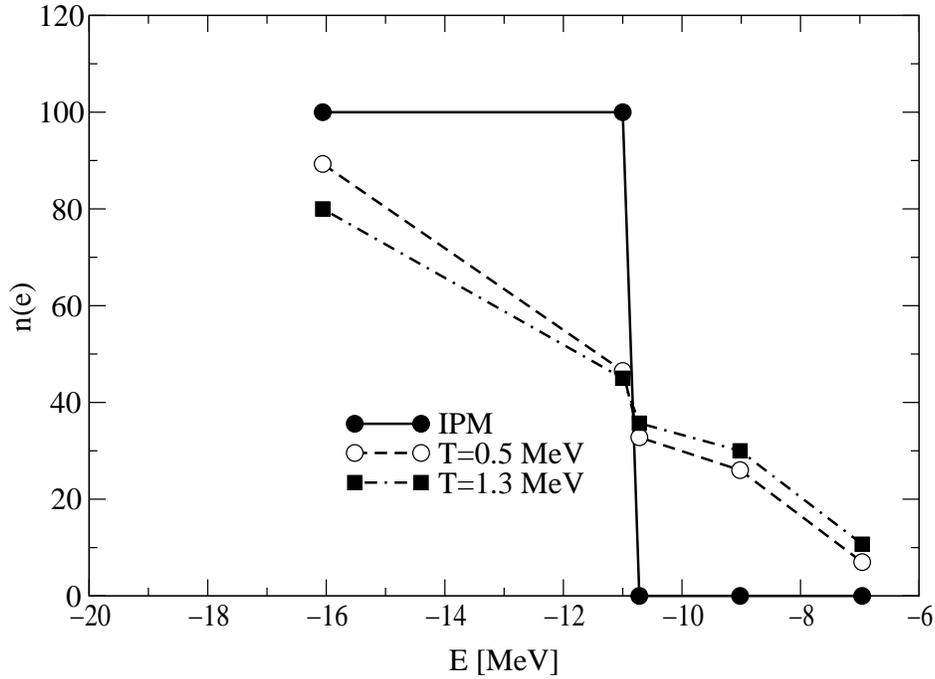}
 \caption{
Proton occupation percentage $n_i(T)/(2j_i+1)$ in $^{74}$Ge for the
independent particle model (IPM, circles) and calculated in the SMMC
approaches at T=0.5 MeV (squares) and 1.3 MeV (diamonds). The single-particle
energies have been determined from a Woods-Saxon potential: -16.1
MeV ($f_{7/2}$), -11.0 MeV ($p_{3/2}$), -10.7 MeV
($f_{5/2}$), -9.0 MeV ($p_{1/2}$), and -7.0 MeV ($g_{9/2}$).
}

\label{figure1a}
   \end{center}
\end{figure}

\begin{figure}[]
   \begin{center}
 \leavevmode
 \includegraphics[width=0.5\columnwidth,angle=270]{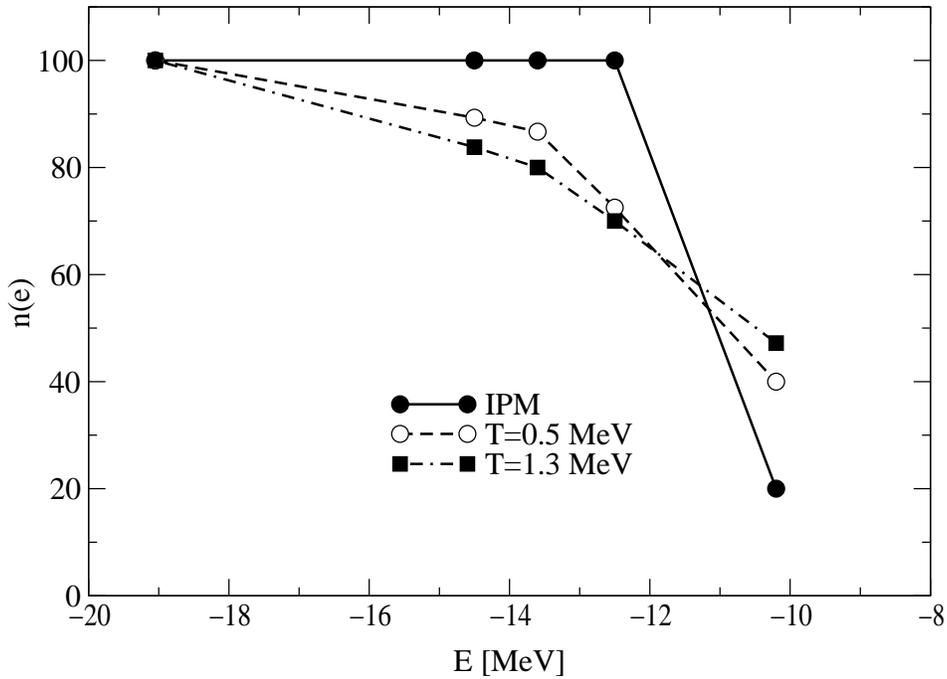}
 \caption{
Neutron occupation percentage $n_i(T)/(2j_i+1)$ in $^{74}$Ge for the
independent particle model (IPM, circles) and calculated in the SMMC
approaches at T=0.5 MeV (squares) and 1.3 MeV (diamonds). The single-particle
energies have been determined from a Woods-Saxon potential: -19.1
MeV ($f_{7/2}$), -14.5 MeV ($p_{3/2}$), -13.6 MeV
($f_{5/2}$), -12.5 MeV ($p_{1/2}$), and -10.2 MeV ($g_{9/2}$).
}

\label{figure1b}
   \end{center}
\end{figure}

We performed SMMC calculations for all even
germanium isotopes $^{68-76}$Ge at various temperatures between T=0.5
MeV and 1.3 MeV. As representative examples, Figs. \ref{figure1a}
and \ref{figure1b} show the
calculated proton and neutron occupation numbers for the nucleus
$^{74}$Ge. In the independent particle model (IPM) the 12 valence protons
completely occupy the $f_{7/2}$ and $p_{3/2}$ orbitals, while the
$p_{1/2}, f_{5/2}$, and $g_{9/2}$ orbitals are empty. (Obviously
IPM predicts the same proton occupation numbers for all
germanium isotopes.)  For the 22 valence neutrons, the IPM predicts a
full $pf$ shell and two neutrons occupying the $g_{9/2}$ orbital.
Clearly, the IPM does not allow GT transitions for $^{74}$Ge.

The residual interaction of the shell model distorts the naive
independent particle picture significantly.
Furthermore, thermal excitation of the nucleus further
perturbs the single-particle occupations.
Both the configuration mixing due to the residual interaction and
the thermal excitations of the many-body system act to smear
the Fermi surface.
This physics is captured
by the SMMC approach.
We observe in Fig. \ref{figure1a}
that for both temperatures the occupation of
the proton $p_{3/2}$ orbital is only about half and that even
the occupation of the proton $f_{7/2}$ orbital is
reduced. The occupation of
the neutron $g_{9/2}$ orbital is nearly doubled,
resulting in 2 neutron holes in the $pf$ shell.

Even at $T=0.5$ MeV, the 
SMMC calculation predicts $\langle n \rangle =0.72$ 
protons in the $g_{9/2}$ orbital,
while for the other isotopes the average
number of protons in this orbital is
0.65 ($^{70}$Ge), 0.59 ($^{72}$Ge), 0.52 ($^{74}$Ge), and
0.48 ($^{76}$Ge). For the same temperature, the SMMC     studies
yield 5.8 neutron holes in the $pf$ shell for $^{68}$Ge
(compared to 4 in the IPM). As expected, the number of neutron holes
is    reduced in the other isotopes; we find 4.3
($^{70}$Ge), 3.1 ($^{72}$Ge), 2.0 ($^{74}$Ge), and
1.1 ($^{76}$Ge). These partial occupations make
possible Gamow-Teller transitions since there
are neutron holes in the $pf$ shell and protons in the $g_{9/2}$
orbital; therefore, we expect from these occupation numbers that
GT transitions for electron capture are not completely blocked. Even
at the rather low temperature $T=0.5$ MeV, configuration
mixing and thermal excitations smear the Fermi surface. This
contrasts with the RPA study of Ref.~\cite{Wambach} 
which found no significant unblocking. The
difference between the
results is that the SMMC
calculations consider the thermal excitation
of all many-body states, while Ref. \cite{Wambach} considered   the
thermal excitation of only the single-particle states in
the model space. The unblocking that we find might
be quite important.  Recent presupernova evolution models 
imply that $Y_e$ is $\sim 0.44$ for
$T=0.5$ MeV \cite{Heger2};
this corresponds roughly to the proton-to-nucleon ratio in
$^{72,74}$Ge.

\begin{figure}[]
   \begin{center}
 \leavevmode
 \includegraphics[width=0.5\columnwidth,angle=90]{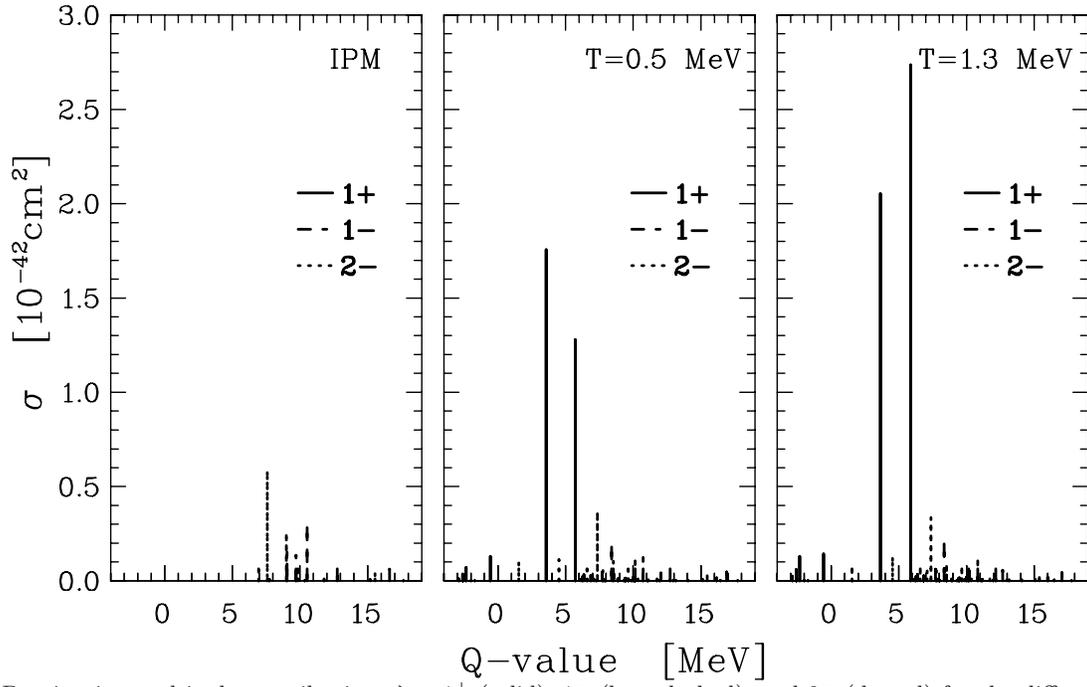}
 \caption{
Dominating multipole contributions
$\lambda =1^+$ (solid), $1^-$
(long-dashed), and $2^-$
(dotted) for the
differential capture cross section for 20 MeV electrons on $^{74}$Ge.
The calculations have been performed within the independent particle
model and for temperatures $T=0.5$ MeV and 1.3 MeV using the hybrid
SMMC/RPA model as described in the text. The Q-value for $T=0$ is
5.4 MeV.}
 \label{figure3}
   \end{center}
\end{figure}

Figure \ref{figure3}
shows the decomposition of the
capture cross section for 20 MeV electrons on $^{74}$Ge into
the leading multipoles
$\lambda =1^+, 1^-, 2^-$.
The calculation has again been performed in the IPM and in the
hybrid model at $T=0.5$ MeV and 1.3 MeV. As noted before
\cite{Fuller82,Wambach}, unblocking is most important for allowed
transitions, although one also observes a slight redistribution of the
strength for the forbidden dipole transitions.
Obviously the differences
are very pronounced for the GT transition. While transitions are blocked
in the independent particle model Gamow--Teller, they dominate the
response in the hybrid model. The two prominent peaks in
the RPA response
correspond to $f_{7/2} \rightarrow f_{5/2}$
and $g_{9/2} \rightarrow g_{7/2}$
proton-neutron transitions.  However,
unblocking also allows particle-to-particle
transitions between partially occupied orbitals. For some of these
transitions, the energy difference between the initial and final
single-particle states is negative and can even lead to
unphysical, negative Q-values (see Fig. \ref{figure3}).
Fortunately these transitions do not
contribute significantly to the total cross section.

Our expectations about the importance of GT unblocking
are realized in the total electron capture calculations
we performed using the RPA method.
The results are exemplified in Fig.\ 4 for
$^{68}$Ge with 4 neutron holes even in the
simple IPM, and for $^{72,76}$Ge for which the IPM does not
allow GT transitions. These differences are well    pronounced in the
IPM capture rates, showing large cross sections for
$^{68}$Ge. For $^{72,76}$Ge, electron capture is mediated by
forbidden transitions resulting in cross sections which are more
than 2 orders of magnitude smaller than for $^{76}$Ge at moderate
electron energies ($E_e < 15$ MeV). Note that capture of an
electron with energy $E_e$ is more difficult on $^{76}$Ge than on
$^{72}$Ge, due to the increased Q-value. We also note that the IPM
cross sections become clearly more similar for larger electron
energies as the sensitivity to the Q-value decreases and   the
relative importance of the forbidden transitions
increases.

When we repeat the calculations for $^{68}$Ge using the occupation numbers
from the SMMC calculations, we find a rather small reduction in
the capture cross section. Since
GT transitions were
already allowed in the IPM, the residual interaction and thermal
effects are quite unimportant for this nucleus.  We also note
that forbidden transitions do not
significantly contribute to electron
capture on $pf$-shell nuclei at moderate electron   energies. This
confirms the assumption made in previous studies that the   capture
rate for $pf$-shell
nuclei during the presupernova evolution can be
solely determined on the basis of GT transitions.

\begin{figure}[]
   \begin{center}
 \leavevmode
 \includegraphics[width=0.5\columnwidth,angle=90]{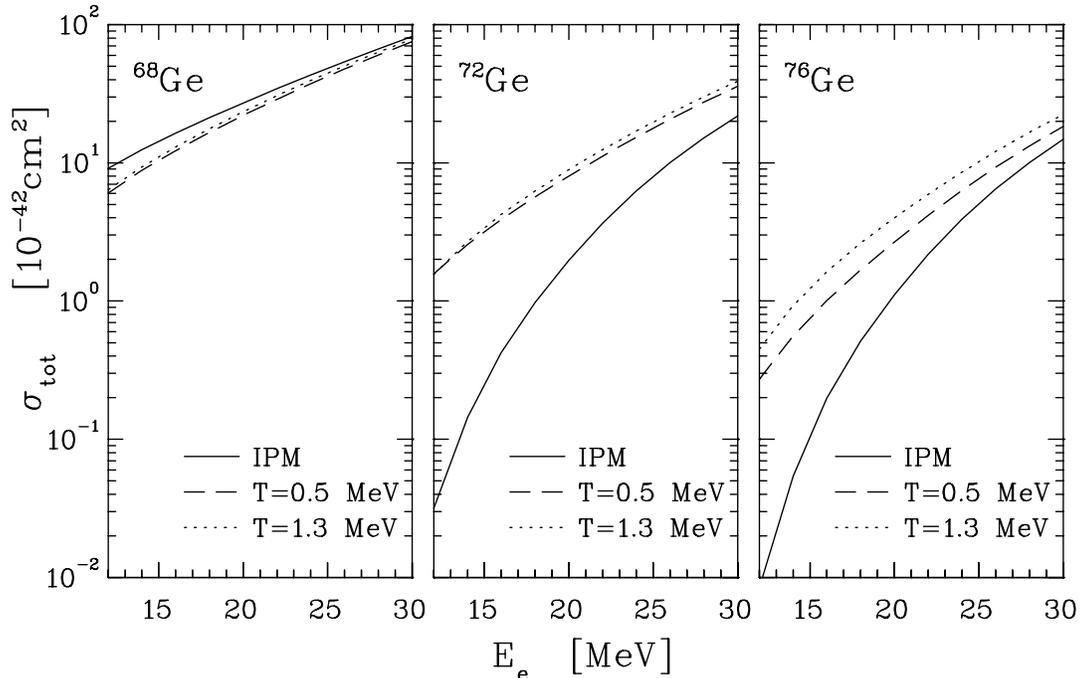}
 \caption{
Electron capture cross sections for
$^{68,72,76}$Ge calculated within the independent particle model (solid)
and for temperatures $T=0.5$ MeV (long-dashed) and 1.3 MeV
(dotted) using the hybrid SMMC/RPA
model as described in the text.}
 \label{figure4}
   \end{center}
\end{figure}

Configuration mixing and thermal excitations
unblock the GT transitions in $^{72}$Ge and
$^{76}$Ge.  This effect is strong enough in both
nuclei to increase the capture cross section by nearly two orders
of magnitude at the lowest electron energies shown in Fig.\ \ref{figure4}.
Our calculations indicate that the unblocking effect is not too sensitive to
increasing temperature.
This is certainly  of strong practical importance as it indicates
that SMMC studies do not have to be performed on very fine temperature
grids, and extrapolations should be quite
sufficient.  At higher electron energies,
forbidden transitions become important since the
differences between the SMMC/RPA
and IPM cross sections diminish.

In summary, in the collapse phase of a supernova, electrons can
be captured on very neutron-rich nuclei with protons in the
$pf$ shell ($Z<40$) and neutron numbers $N>40$. For
these nuclei, Gamow-Teller  transitions, which dominate electron
capture on $pf$-shell nuclei during   the presupernova evolution, are
forbidden in the independent particle model. We have
shown that this model is too
simple for applications to neutron-rich isotopes,             since
GT transitions are unblocked by finite temperature excitations
and by the mixing of occupations
of the $pf$ and $g_{9/2}$
orbitals induced by the residual interaction. In this paper, we
propose a hybrid model in which the temperature
and configuration-mixing effects are studied within the Shell
Model Monte Carlo approach and are described by partial occupation
numbers for the various single-particle orbits. Using the
mean-field wave function with these corresponding
partial occupancies, the electron capture
cross sections are calculated with
an RPA approach. We considered both allowed GT and
forbidden transitions.

We applied our hybrid model to the even
germanium isotopes $^{68-76}$Ge at typical collapse temperatures
$T \sim 0.5-1.5$ MeV.   At all temperatures,
the residual interaction is sufficiently
strong to unblock the GT transitions which then also
dominate stellar electron capture on these nuclei. However, with
increasing electron energies, i.e., at larger electron
chemical potentials and temperatures, forbidden transitions
become increasingly important and can no longer be neglected.
The present model  consistently describes allowed and
forbidden transitions and should thus be also
applicable to such situations.

This work was supported in part
by the Danish Research Council.
Oak Ridge National Laboratory is managed
by UT-Battelle, LLC under
Contract No.  DE-AC05-00OR22725 for the U.S.
Department of Energy.

\end{document}